\documentclass[showpacs,floatfix,twocolumn,prl,superscriptaddress]{revtex4-1}

\usepackage{graphicx,color}

\newcommand*{\kfsfive}{K$_2$Fe$_4$Se$_5$}
\newcommand*{\kfstwo}{K$_{0.8}$Fe$_{1.6}$Se$_2$}
\newcommand*{\kfssc}{K$_{0.85}$Fe$_{1.9}$Se$_2$}
\newcommand*{\kfsmetal}{K$_x$Fe$_2$Se$_2$}
\newcommand*{\kfsxy}{K$_x$Fe$_{2-y}$Se$_2$}
\newcommand*{\kfsfull}{KFe$_2$Se$_2$}

\begin{document}

\title{Phase relations in K$_x$Fe$_{2-y}$Se$_2$ and the structure of superconducting K$_x$Fe$_2$Se$_2$ via high-resolution synchrotron diffraction}

\author{Daniel P. Shoemaker}
\author{Duck Young Chung}
\author{Helmut Claus}
\author{Melanie C. Francisco}
\author{Sevda Avci}
\affiliation{Materials Science Division, Argonne National Laboratory,
Argonne, IL, 60439, USA}

\author{Anna Llobet}
\affiliation{Los Alamos National Laboratory, Lujan Neutron Scattering Center, 
MS H805, Los Alamos, New Mexico 87545, USA}

\author{Mercouri G. Kanatzidis}\email{m-kanatzidis@northwestern.edu}
\affiliation{Materials Science Division, Argonne National Laboratory,
Argonne, IL, 60439, USA}
\affiliation{Department of Chemistry, Northwestern 
University, Evanston, Illinois 60208, United States}


\begin{abstract}
Superconductivity in iron selenides has experienced a rapid growth,
but not without major inconsistencies in the reported properties. For
alkali-intercalated iron selenides, even the structure of the superconducting
phase is a subject of debate, in part because the onset of superconductivity
is affected much more delicately by stoichiometry and preparation than
in cuprate or pnictide superconductors.
If high-quality, pure, superconducting
intercalated iron selenides are ever to be made, the intertwined physics and chemistry
must be explained by systematic studies of how these materials form and
by and identifying the many coexisting phases.  To that end, we prepared
pure K$_2$Fe$_4$Se$_5$ powder and superconductors in the K$_x$Fe$_{2-y}$Se$_2$ system, and 
examined differences in their structures by high-resolution synchrotron and 
single-crystal x-ray diffraction. We found
four distinct phases: semiconducting K$_2$Fe$_4$Se$_5$, a
metallic superconducting phase K$_x$Fe$_2$Se$_2$ with $x$ ranging from
0.38 to 0.58,
an insulator KFe$_{1.6}$Se$_2$ with no vacancy ordering,
and an oxidized phase K$_{0.51(5)}$Fe$_{0.70(2)}$Se
that forms the PbClF structure upon exposure to moisture.
We find
that the vacancy-ordered phase K$_2$Fe$_4$Se$_5$ does not become superconducting
by doping, but the distinct iron-rich minority phase K$_x$Fe$_2$Se$_2$ precipitates from
single crystals upon cooling from above the vacancy ordering temperature.
This coexistence of metallic and semiconducting phases explains
a broad maximum in resistivity around 100 K.
Further studies to understand the solubility of excess Fe in the K$_x$Fe$_{2-y}$Se$_2$
structure will shed light on the maximum fraction of superconducting K$_x$Fe$_2$Se$_2$
that can be obtained by solid state synthesis.
\end{abstract}


\maketitle

\section{Introduction} 

The brief history of iron chalcogenide superconductivity has seen
a flurry of activity, beginning with the
discovery of $T_c$ = 8 K in $\beta$-FeSe,\cite{hsu_superconductivity_2008}
 and later the announcement that ternary intercalated
compounds in the $A_x$Fe$_{2-y}$Se$_2$ system display $T_c \approx 30$ K
when $A$ is K, Rb, Cs, or Tl. 
\cite{guo_superconductivity_2010, ye_common_2011}.
Much like the superconducting iron arsenides, these compounds form the
ThCr$_2$Si$_2$ structure-type with layers 
of tetrahedrally-coordinated
Fe and are in the vicinity of antiferromagnetism, but the differing 
anion charges (formally Se$^{2-}$ versus As$^{3-}$) lead to issues
of chemical stability that have a profound effect on the structures
and properties.
While arsenides are only known to exhibit superconductivity in the
fully-occupied ThCr$_2$Si$_2$ structure type without vacancies,
the hallmark of the selenides (intercalated and not) is that stoichiometry is never
exact for superconducting samples---some disorder is 
always present, often in conjunction with phase separation.
\cite{wen_interplay_2011,ivanovskii_new_2011,mou_structural_2011,syu_synthesis_2011}

Experimental efforts to understand superconductivity in iron selenides 
must grapple with the sensitive stoichiometry
required to observe $T_c$. Compared to iron pnictides, where
a superconducting dome appears from $x$ = 0.2 to 1  
in Ba$_{1-x}$K$_x$Fe$_2$As$_2$ for example, \cite{avci_phase_2012}
in $\beta$-Fe$_{1+\delta}$Se there is only a 
window of $\delta$ = 0.01 to 0.03 where superconductivity is observed, 
and there is no
such dome versus composition.\cite{mcqueen_extreme_2009}
No dome is present in $A_x$Fe$_{2-y}$Se$_2$ superconductors either,
with $T_c$ approximately invariant around 30 K.
\cite{yan_electronic_2012,fang_fe-based_2011,bao_vacancy_2011}
Additionally, the thermal
history of the sample plays a key role, as even moderate thermal annealing
has an effect on the sharpness of the transition in $A_x$Fe$_{2-y}$Se$_2$.
\cite{han_metastable_2012,ozaki_one-step_2012}

\begin{figure}
\centering\includegraphics[width=0.8\columnwidth]{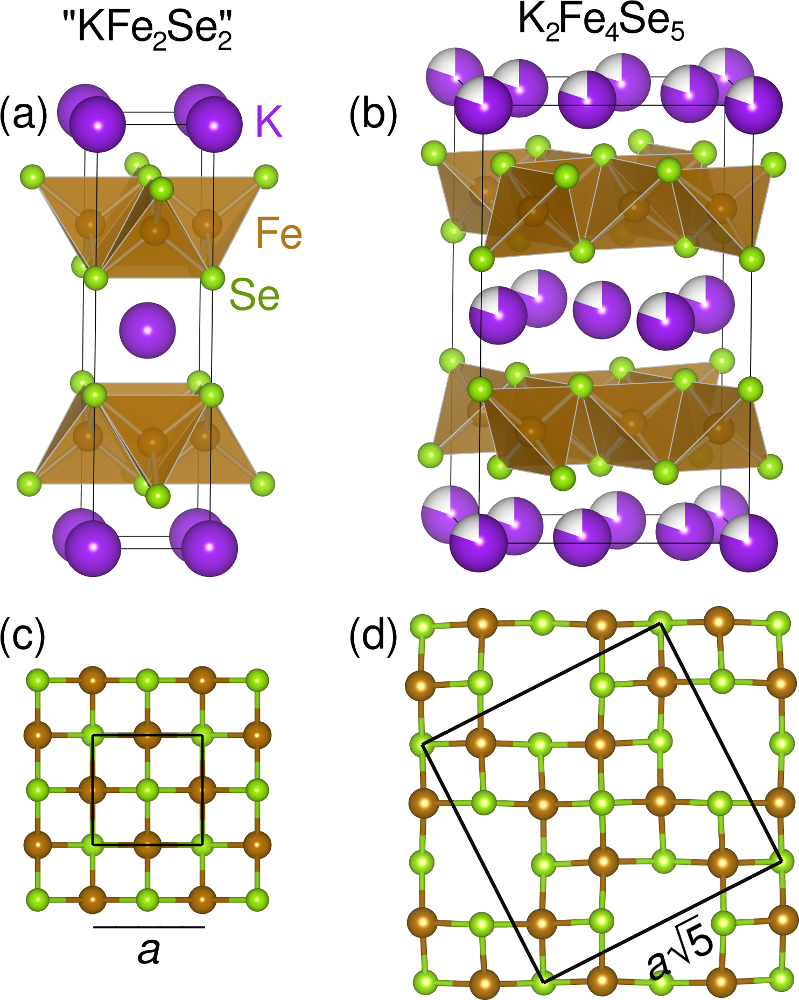} \\
\caption{(Color online) Unit cells for (a) hypothetical $I4/mmm$ fully-occupied \kfsfull\
and (b) Fe-vacancy ordered \kfsfive\, which is equivalent to \kfstwo. Fe/Se nets are
viewed down the $c$ direction in (c) and (d).
}
\label{fig:unitcells}
\end{figure}

Divalent Se$^{2-}$ leads to the presence of alkali and iron vacancies
that are not found in the iron arsenides. In fact, most attention in 
the \kfsxy{} system is focused on \kfstwo{}, shown
in Figure \ref{fig:unitcells}. This compound is
a Mott insulator with 1/5 ordered
Fe vacancies and disordered K, and can be written as \kfsfive{} with
valence-precise Fe$^{2+}$. \cite{bacsa_cation_2011} 
Due to prevalent vacancies and
the ability of Fe to adopt +2 or +3 
formal oxidation states, 
(as in Fe$^{2+}$Se or KFe$^{3+}$Se$_2$ \cite{bronger_antiferromagnetic_1987})
it may seem that doping either
cation in \kfsxy{} would tune $T_c$ as in the arsenides, but this
is not the case: the superconducting transition appears
and disappears abruptly, and does not shift.\cite{yan_electronic_2012}

A synthetic route to pure superconducting \kfsxy{} phases is elusive.
Microscale phase separation between closely-related structures,
mobile Fe/K vacancies, and iron impurities are widespread.
\cite{guo_superconductivity_2010,ryan_fe_2011,torchetti_se_2011, ksenofontov_phase_2011, ma_local_2011, yan_ternary_2011,lv_vacancy_2011}
As a result, models or measurements that describe properties without accounting
for sample heterogeneity are up to now incomplete.
For example, photoemission spectroscopy, 
energy dispersive spectroscopy, and inductively-coupled plasma
spectroscopy can only probe the composition of large portions
of the samples--for a heterogeneous sample they do not describe 
any single component.
Resolving multiple phases simultaneously is key in these systems,
where a second metallic phase apart from pure \kfsfive{} is believed
to lead to superconductivity on the basis of NMR, 
muon spin resonance, and scanning probe
measurements.\cite{torchetti_se_2011,texier_nmr_2012,charnukha_nanoscale_2012,chen_electronic_2011}

In order to understand why some samples are superconducting and some are not,
we have conducted a systematic investigation of many samples. 
We prepared pure \kfsfive{} and verified
its existence using high-resolution synchrotron x-ray diffraction.
We prepared
superconducting crystals and investigated the changes in the \kfsxy{} lattice,
including the appearance of three distinct additional phases: the metallic
\kfsmetal{}
phase that precipitates coherently with \kfsfive{} upon cooling
and is the cause of superconductivity,
a PbClF-type phase that forms due to exposure
to moist air, and the insulator KFe$_{1.6}$Se$_2$ with full K occupancy
and disordered Fe vacancies.
All of these phases must be understood and controlled in order to explain
the properties and diffraction data.
We also show that the anomalous resistivity behavior, previously
thought to signify a metal-insulator transition, 
\cite{lei_phase_2011} in fact arises from simple 
percolation of metallic and insulating phase fractions. With a more
complete picture of the phase space in the \kfsxy{} system, we discuss
implications for improved synthetic routes to superconducting
intercalated iron selenides.

\section{Methods}

Samples of \kfsxy\ were prepared from metallic K,
Fe powder, and crushed Se shot (Alrich, 99.5\%, 99.99\%, and 99.99\%, respectively). 
All manipulations were performed in a N$_2$-filled glovebox.
Stoichiometric powders, including pure \kfsfive, were prepared
by intimately mixing Fe and Se in a mortar and pestle
in a N$_2$-filled glovebox with a ratio of 4Fe + 5Se,
then loading in a carbon-coated quartz tube and sealing under vacuum.
This tube was heated with a 12 h ramp to 700$^\circ$C, 2 h hold, and
furnace cool back to room temperature. This powder was ground again
in a glovebox and loaded with K pieces in a covered alumina crucible
in a quartz tube, sealed under vacuum, and heated over the same temperature
profile. Finally, the powder was homogenized by grinding and fired
with a 1 h ramp to 700$^\circ$C, 10 h hold, and 1 h cool to room
temperature.

Single crystals were prepared by prereaction of K pieces with
Fe and Se powder in alumina crucibles
sealed under vacuum and heated to 600 or 650$^\circ$C in 12 h, with a 4 h hold and
4 h cool to room temperature, followed by grinding. Slow-cooled
crystal growth was performed in alumina crucibles sealed under Ar in Nb tubes.
Flame-melted samples were prepared by melting the prereacted powders
in evcuated quartz tubes until the mixture was visibly molten.
The nominal composition \kfsxy{} was varied from $0.8 \leq x \leq 0.85$
and $0 \leq y \leq 0.4$.
Specific compositions and heat treatments are presented in the Supplemental
Material.

High-resolution ($\Delta Q/Q < 2\times10^{-4}$) synchrotron powder diffraction data were collected using beamline
11-BM at the Advanced Photon Source (APS), Argonne National Laboratory using an
average wavelength of 0.413 \AA\ ($\sim$30 keV). 
A NIST standard Si sample (SRM 640c) was used to calibrate the
instrument, where the Si lattice constant determines the wavelength for each detector.
Samples were sealed under vacuum in glass capillaries to prevent oxidation.
Time-of-flight powder neutron diffraction measurements were conducted at the
HIPD instrument at the Lujan Center, Los Alamos National Laboratory with
samples sealed under He in vanadium cans. Rietveld refinements to
synchrotron x-ray and neutron diffraction data were performed using 
GSAS.\cite{larson_general_2000}

Laboratory x-ray powder diffraction was performed using a 
Philips X'Pert diffractometer with Cu-$K\alpha$ radiation, and
Rietveld refinements were performed using the XND code.\cite{berar_xnd_1998}
Single-crystal diffraction data were collected on a STOE 2T image plate 
diffractometer with Mo-$K\alpha$ radiation ($\lambda$ = 0.71073 \AA) 
and X-Area software, and structures were refined using SHELXTL.\cite{sheldrick_short_2007}
Four-probe resistivity, ac magnetic susceptibility, and heat capacity were
measured using a Quantum Design PPMS.

\section{Results and Discussion}

\subsection{Characterization of pure, polycrystalline \kfsfive}
\begin{figure}
\centering\includegraphics[width=0.85\columnwidth]{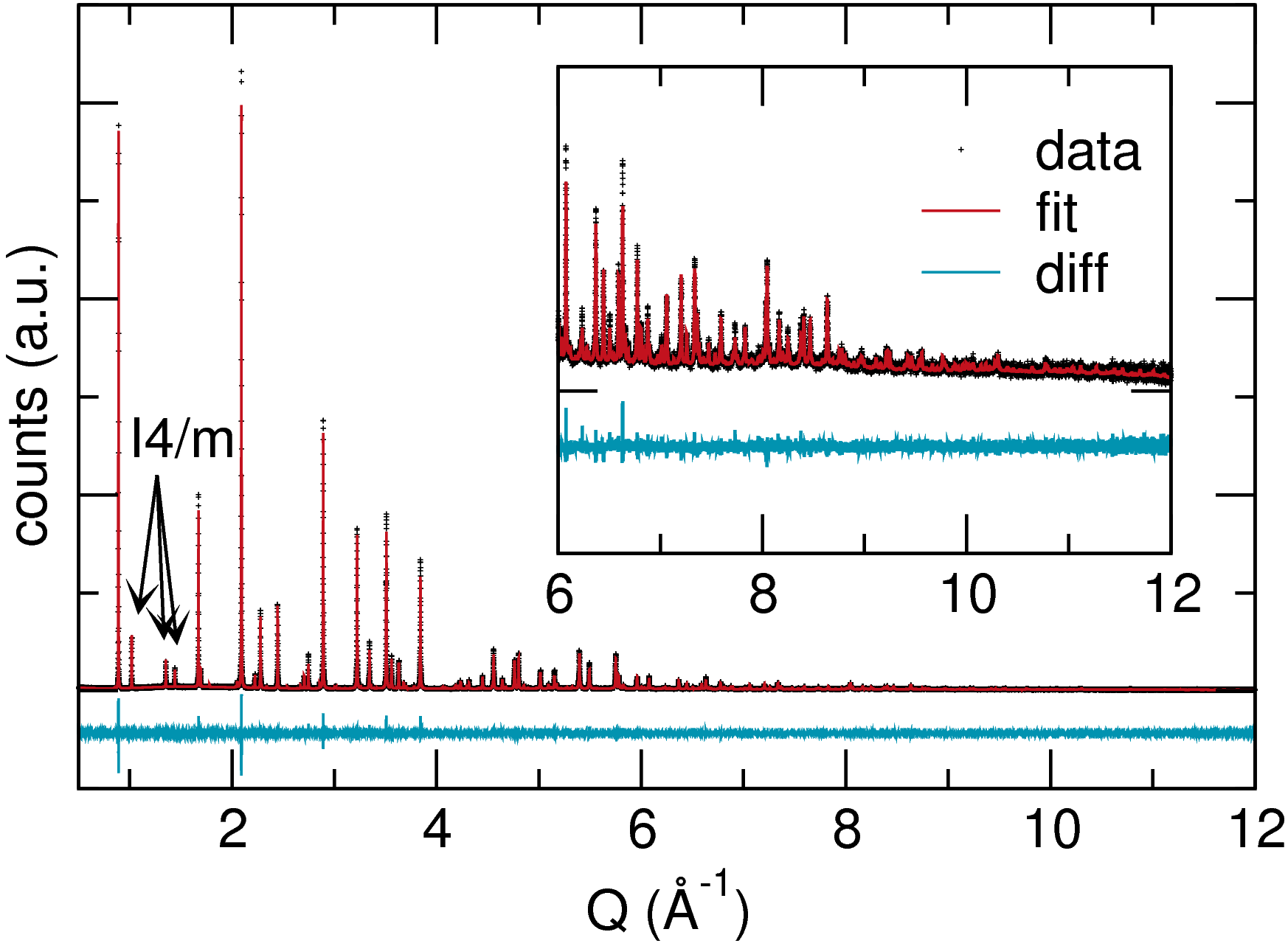} \\
\caption{(Color online) Rietveld refinement to high-resolution synchrotron
diffraction data for 
powder \kfsfive{} shows a pure compound with nearly complete vacancy ordering:
only 7\% of the Fe $4d$ sites are occupied. Low-angle peaks corresponding to the $I4/m$
cell due to Fe ordering are arrowed. High-$Q$ data are enlarged in
the inset to show fit quality. Structural details are given in Supplementary 
Information.
}
\label{fig:rietveld-245}
\end{figure}

The composition-temperature phase space of \kfsxy{} is mostly unknown, so the need for
a pure, homogeneous sample that can serve as a reference point is
paramount. The most stable phase near superconductivity
in this family is vacancy-ordered
\kfsfive{} with the unit cell shown in Figure \ref{fig:unitcells}. In this
structure, first reported in TlFe$_{1.6}$Se$_2$,\cite{haggstrom_magnetic_1986} the Fe vacancies order (lowering symmetry from
$I4/mmm$ to $I4/m$) but the K vacancies are distributed randomly. 
\cite{bacsa_cation_2011} Single crystals of this compound can be grown from
the melt,
but high-temperature processing that involves melting results in
samples that deviate from nominal stoichiometry

We developed a lower-temperature,
solid-state route to form pure \kfsfive{}.
The low-temperature procedure described in the experimental section
consists of a prereaction of Fe and Se, followed
by addition of K and multiple heatings to 700$^\circ$C.
To confirm phase purity we performed powder
diffraction at APS beamline 11-BM, which provides exceptionally high-resolution
data with high signal-to-noise ratio, while also maintaining capillary geometry
that prevents air exposure. The Rietveld refinement shown in Figure \ref{fig:rietveld-245}
consists of sharp, unsplit peaks with no impurity phases, confirming the sample
quality and homogeneity.
At low angles the superstructure peaks from vacancy ordering are clearly
visible, and arrowed in Figure \ref{fig:rietveld-245}.
This sample refines to nearly complete vacancy ordering:
only 7\% of the Fe $4d$ sites are occupied. Detailed
refinement results are given in the Supplemental Material.

This is a simple, reliable method for producing pure \kfsfive.
Our magnetometry
and resistivity measurements confirmed that \kfsfive{} is an antiferromagnetic
semiconductor.\cite{bacsa_cation_2011,song_phase_2011,svitlyk_temperature_2011} While
this powder synthesis provides great compositional control, we have never
observed superconductivity in any powders created by this method, even when
changing the stoichiometry in \kfsxy\, where $0.5 < x < 1$ and $1.4 < y < 2$.

This stoichiometric polycrystalline powder sample is crucial because it sets
a structural reference point
for which all other compositions will be compared. There is no evidence
(line broadening, extra peaks, extra phases) in the 11-BM diffraction data
for phase separation when pure \kfsfive{} is made by this route.

\subsection{Structural characterization of nominal \kfsfive{} crystals}

To date, there has been no mention of a superconducting powder of 
\kfsxy{}, nor did we find one despite our efforts. 
This implies that melting and recrystallization may be required for the 
formation of the superconducting phase.
We prepared single crystals of nominal \kfsfive{} composition
to determine how stoichiometry is affected by melting. Crystals
prepared by melting nominal \kfsfive{} formed plates
which readily degrade in air, as judged by a change in color from
shiny gold to matte brown. 

\begin{figure}
\centering\includegraphics[width=0.85\columnwidth]{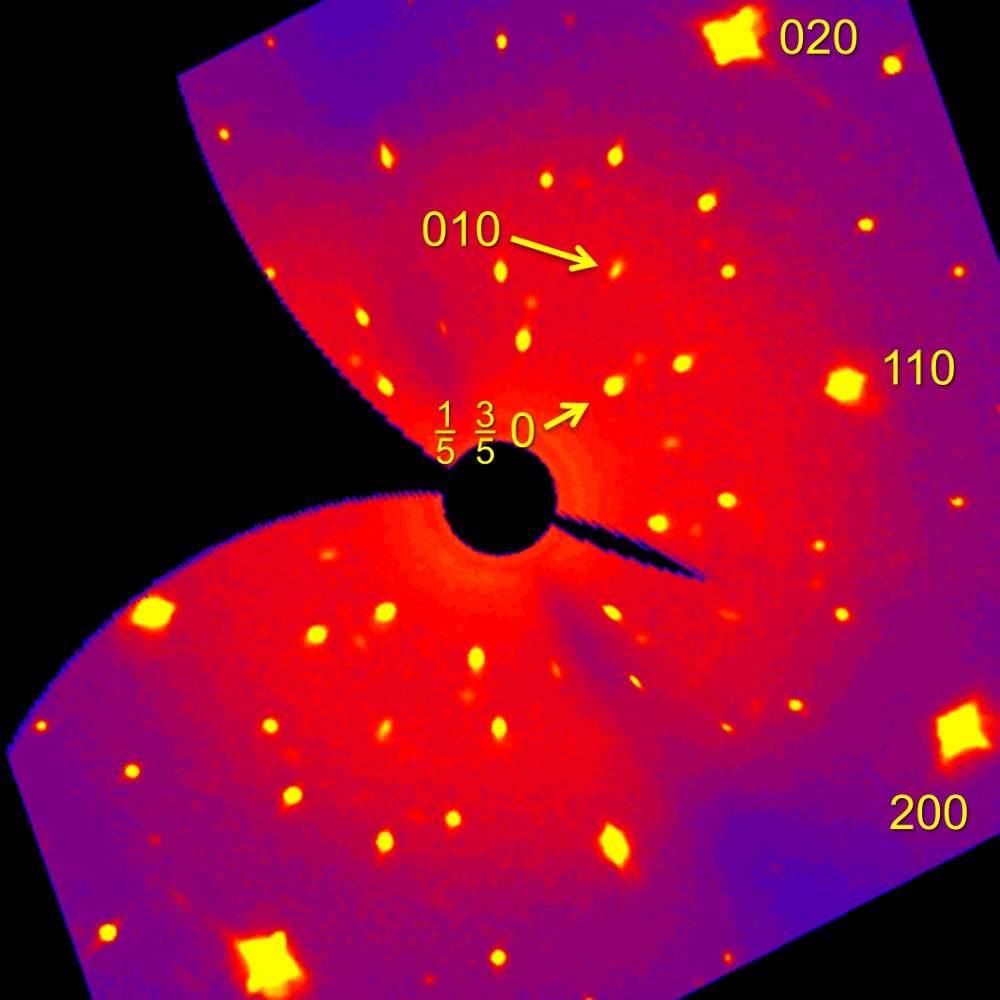} \\
\caption{(Color online) Reciprocal space reconstruction of 
single crystal x-ray diffraction data from a nominal \kfsfive{} crystal.
Reflections are labeled with Miller indices of the $I4/mmm$ ThCr$_2$Si$_2$
substructure. The fractional superstructure peaks, including
the labeled peak at ($\frac{1}{5} \frac{3}{5} 0$), arise
from vacancy ordering and lowering of symmetry to $I4/m$
The (010) reflection is forbidden by both $I$-centered cells, and 
represents a new, coherent phase. Subsequent analysis
in this manuscript confirms it to be an oxidized phase with $c$ = 9 \AA.
}
\label{fig:kspace-245crystal}
\end{figure}

Single crystal diffraction of these nominal \kfsfive{} crystals
shows superstructure Bragg
peaks arising from $I4/m$ \kfsfive{}.
These peaks form an octagon in the $(00l)$ reciprocal-space reconstruction
in Figure \ref{fig:kspace-245crystal}, with the first peak at $(\frac{1}{5} \frac{3}{5} 0)$
arrowed. Extra reflections 
appear at the (010) position of the $I4/mmm$ \kfstwo{} lattice (arrowed
in Figure \ref{fig:kspace-245crystal}) which
is forbidden by $I$-centered symmetry. 
They do not represent a $\sqrt 2 \times \sqrt 2$ modification of the \kfsxy{} structure,
but instead arise from an oxidized phase that will be discussed in the next section.
No other vacancy ordering patterns are observed in these crystals.

\begin{figure}
\centering\includegraphics[width=0.85\columnwidth]{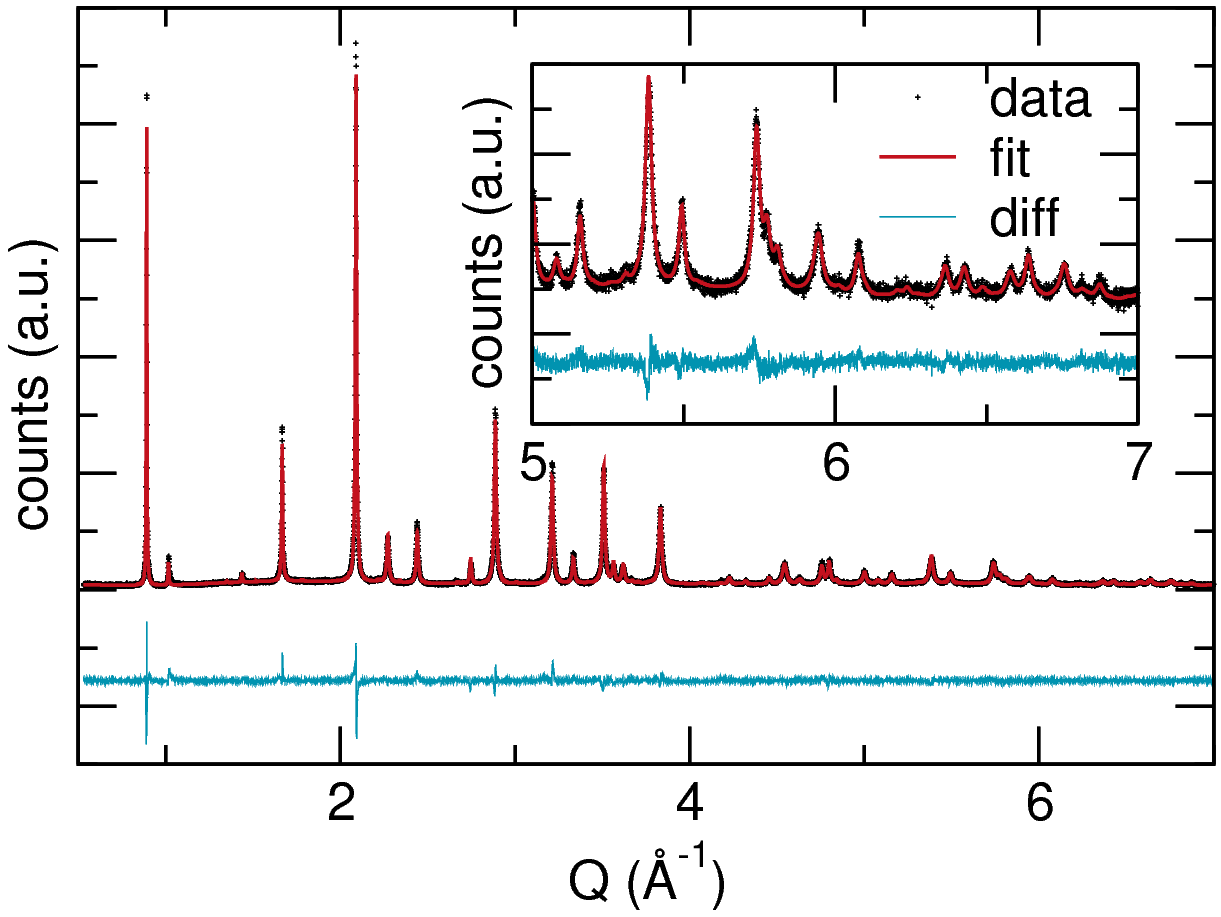} \\
\caption{(Color online) Rietveld refinement to high-resolution synchrotron
diffraction data for a non-superconducting, ground single-crystal sample
of nominal \kfsfive{} composition. This sample displays $I4/m$ vacancy
ordering.
}
\label{fig:rietveld-245crystal}
\end{figure}

High-resolution x-ray diffraction was performed on ground batches of these crystals
to search for phase separation in the form of split c-axis reflections,
seen often in superconducting samples
\cite{luo_crystal_2011,ozaki_one-step_2012,wang_superconductivity_2011}
and to screen for any minor impurities.
Both are absent, and the fit from Rietveld refinement
is shown in Figure \ref{fig:rietveld-245crystal}.

We found that the lattice constants of nominal \kfsfive{} crystals
display an expanded $a$ and contracted $c$-axis compared to the
pure powder \kfsfive{} 
(8.74536(8) $\times$ 14.10024(18) \AA{} versus 
8.721763(10) $\times$ 14.125178(23) \AA{} for powder \kfsfive{}).
The refined stoichiometry of the crystal was K$_{0.79(1)}$Fe$_{1.56(1)}$Se$_2$,
while the ground batch of crystals had a refined composition
of K$_{0.84(1)}$Fe$_{1.43(1)}$Se$_2$ from synchrotron powder diffraction.
The difference between powder and single-crystal measurements likely
arises from heterogeneity among the crystals or systematic
errors, but in any case 
both techniques find that \kfsfive{} becomes Fe deficient after
melting and recrystallization, and does not exhibit superconductivity.
Still, the presence of an impurity phase in the single crystals
merits further investigation, primarily to understand and avoid
its conditions for formation.

\subsection{Forbidden (010) diffraction spots arise from
the oxidized phase K$_{0.51(5)}$Fe$_{0.70(2)}$Se}

\begin{figure}
\centering\includegraphics[width=0.8\columnwidth]{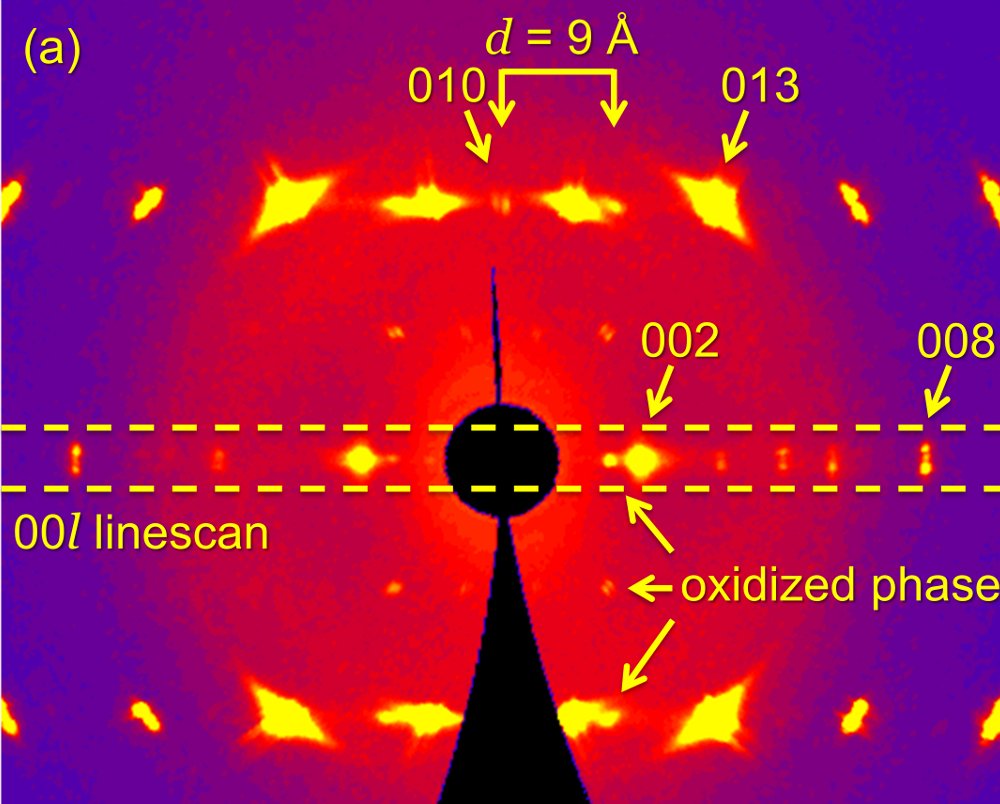} \\
\centering\includegraphics[width=0.85\columnwidth]{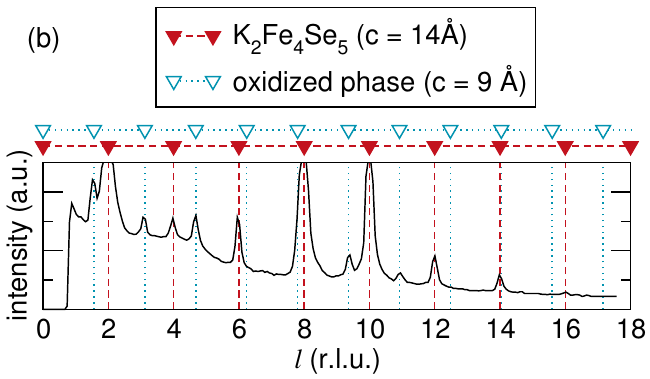} \\
\caption{(Color online) 
The reciprocal space reconstruction along the (100)
direction of the parent $I4/mmm$ \kfstwo{} lattice shows that the extra
reflections, including (010) from Figure \ref{fig:kspace-245crystal},
lie along an $l$ index that is distinct from \kfstwo. This distinct
spacing is shown with $d$ = 9 \AA. An intensity linescan
along $00l$ in (b) shows that these spots arise from a phase where
the FeSe interlayer spacing is 9 \AA, as opposed to $c/2$ = 7 \AA{} for 
\kfsfive.
}
\label{fig:oxidized-frame}
\end{figure}

The extra (010) Bragg reflections in the 
\kfstwo{} reciprocal space reconstruction in Figure \ref{fig:kspace-245crystal}
merit further investigation to understand whether they might correspond
to a $\sqrt 2 \times \sqrt 2$ superstructure of the \kfsxy{} cell.
Such a cell has been
proposed on the basis of electron diffraction patterns 
viewed down the $\langle 001 \rangle$ direction.\,\cite{kazakov_uniform_2011,li_collapse_2011}
No such phase has ever been
made in bulk quantities or detected by x-ray diffraction, and
the electron diffraction peaks were not shown in the ($0kl$) or ($h0l$) directions
to confirm registry with the \kfsxy{} lattice.

In Figure \ref{fig:oxidized-frame} we present the single crystal 
diffraction pattern 
from a perpendicular direction, down $\langle 100 \rangle$ 
in the ($0kl$) plane. From
this vantage point the extra reflections form a vertical column with
an $l$-spacing that is distinct from
the major \kfstwo{} peaks in the diffraction pattern. This column
is at a distance $d$ = 9 \AA$^{-1}$ from $l$ = 0, arrowed in Figure \ref{fig:oxidized-frame}(a).
A line scan along the $\langle 00l \rangle$ direction
produced the intensity profile in Figure \ref{fig:oxidized-frame}(b).
The $00l$ reflections  for the major \kfstwo{} peaks are marked by
dashed lines, while the minority phase is dotted.
Assuming a tetragonal structure still built of FeSe tetrahedral layers, the smaller 
reciprocal-space repeat distance of the minor phase corresponds
to an interlayer spacing of $d$ = 9 \AA.

\begin{figure}
\centering\includegraphics[width=0.85\columnwidth]{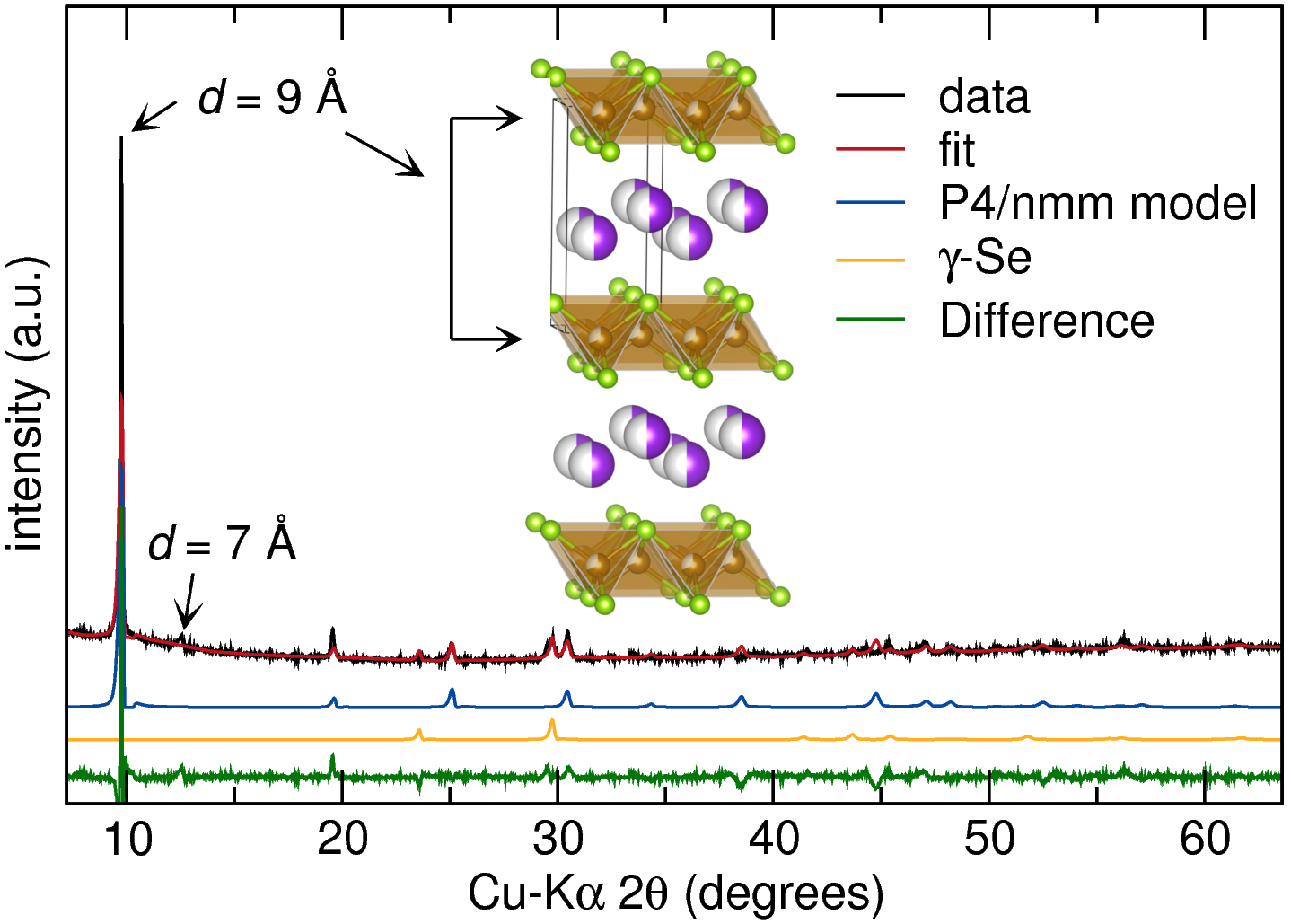} \\
\caption{(Color online) Powder XRD of the \kfsfive{} after exposure
to moist air shows conversion to the oxidized phase K$_{0.51(5)}$Fe$_{0.70(2)}$Se
with the PbClF structure, containing buckled K$^+$ layers and
Fe vacancies. Results from single-crystal structure solution
are given in the Supplemental Material.}
\label{fig:oxidized-xrd-cell}
\end{figure}

\begin{table}
\caption{\label{tab:oxidized} 
Single-crystal refinement results for the oxidized phase
K$_{0.51(5)}$Fe$_{0.698(19)}$Se. Space group: $P4/nmm$, $a$ =  3.8952(6) \AA, $c$ = 9.1948(18)\AA.
Full refinement details are given in the Supplemental Material.
}
\centering
\begin{tabular}{llllllll}
\hline
Atom		& x		& y		& z				& $U_{11}=U_{22}$	& $U_{33}$ (\AA$^2$)	& occupancy \\
K		$2c$	& 0.75	& 0.75	& 0.428(4)	& 0.140(10)						& 0.150(20)				& 0.51(5) \\
Fe		$2a$	& 0.75	& 0.25	& 0				& 0.045(3)							& 0.120(8)				& 0.698(19) \\
Se		$2c$	& 0.25	& 0.25	& 0.1559(6)	& 0.061(2)							& 0.116(4)				& 1 \\
\hline
\end{tabular}
~\\
\end{table}

In our case, this new phase is formed when crystals are
screened and mounted for single crystal diffraction in paratone oil.
Once the crystals are selected and placed in capillaries, the tubes are
sealed and oxidation halts, resulting in only a minor fraction of this oxidized
phase. If powder is exposed to moisture (dry oxygen does not react) for prolonged
periods, full conversion to the oxidized phase occurs, as shown
in the powder diffraction pattern in Figure \ref{fig:oxidized-xrd-cell}.
After full conversion to the new phase the supercell ordering disappears
but the $c$ = 9 \AA{} Bragg peak remains.
Only a tiny peak remains at $2 \theta$ = 13$^\circ$, indicating
almost full degradation of \kfsxy{}. The new phase was determined from
single-crystal diffraction to be a highly K- and Fe- deficient 
structure of the PbClF structure type, K$_{0.51(5)}$Fe$_{0.70(2)}$Se.
This structure is common to a wide range of compounds, including the
superconductor NaFeAs, which itself transforms to a ThCr$_2$Si$_2$ structure
upon hydration.\cite{todorov_topotactic_2010}
Results from single-crystal refinement are shown in Table \ref{tab:oxidized},
with full details given in the Supplemental Material.
No superconducting behavior was seen in any samples
after conversion to the oxidized phase. 

Care must be taken
to avoid air exposure of these samples, especially when surface-sensitive
measurements are made.
The expulsion of Se from the structure seen in Figure \ref{fig:oxidized-xrd-cell}
implies that Fe$^{2+}$ is being oxidized. This phase 
may explain why substantial $c$-axis disorder was seen in
x-ray absorption measurements.\cite{iadecola_large_2012}
Abnormally small Fe--Fe distances
were also seen in that study, which can be explained by the 
metallic superconducting minor phase which we address subsequently.

Formation of this rapidly forming oxidized phase may go unobserved 
in surface-sensitive measurements since it
is coherent with the parent phase of \kfsfive{} from which it originates. Therefore, 
any studies of these samples where a significant exposure to air
has occurred during handling (several minutes) may be tainted by interference from the
oxidized phase of K$_{0.51(5)}$Fe$_{0.70(2)}$Se.
Presence of this oxidized phase and the superconducting minority phase,
which we discuss subsequently, should be considered when interpreting
angle-resolved photoemission spectroscopy in particular, where
a $\sqrt 5 \times \sqrt 5$ supercell is not seen.\cite{berlijn_effective_2012}

\subsection{Superconducting \kfsxy: Changes in the $I4/m$ majority
phase and evidence for phase separation}

We grew superconducting crystals using the same procedure
as our nominal \kfsfive\ crystals, except the nominal stoichiometry was 
\kfssc.
These crystals appear visually similar,
but excess iron precipitates as metal and often pervades the solidified
ingot, with its highest concentration at the top of the ingot.
Iron is denser than \kfsfive{} (7.8 versus 4.3 g/cc) so it was
most likely pushed upward by the advancing solidification front
and not floating on the selenide melt. The extent of Fe solubility
in \kfsxy{} melts remains unknown, and may be a key in determining
how to control phase separation in these materials.

\begin{figure}
\centering\includegraphics[width=0.85\columnwidth]{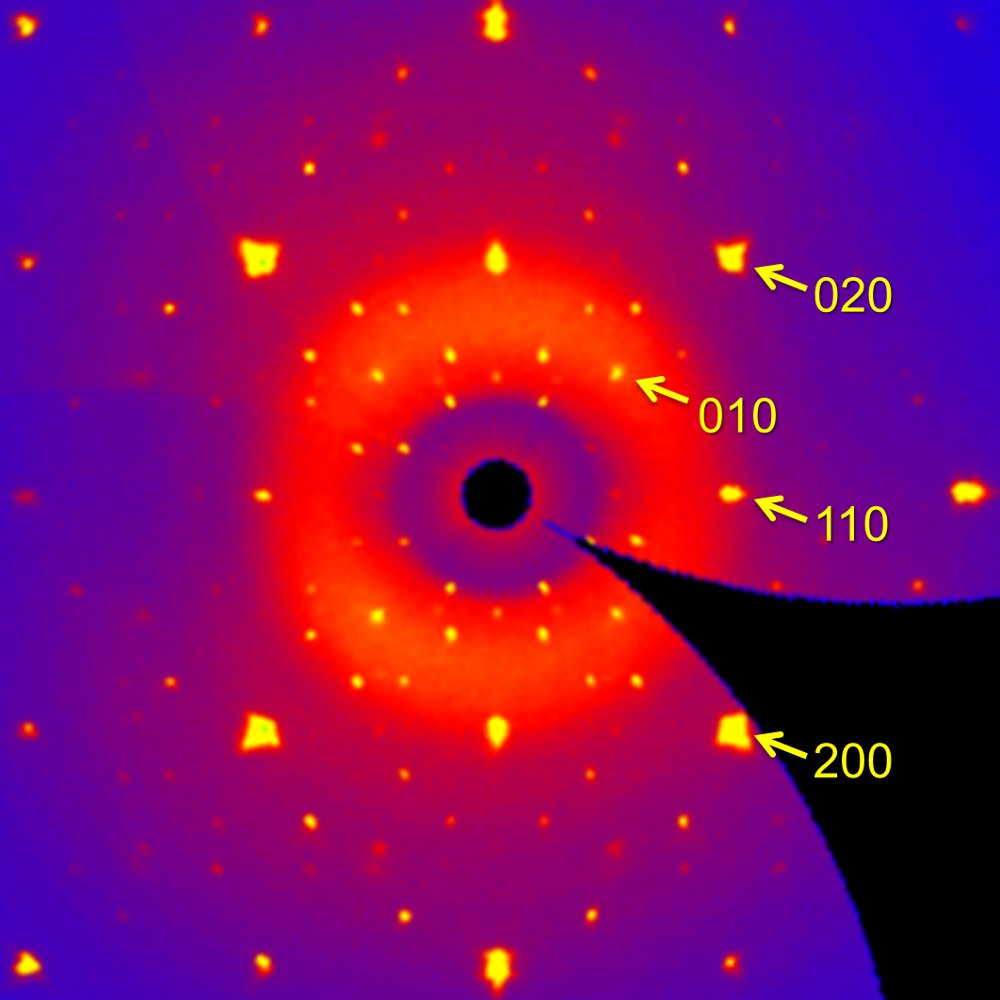} \\
\caption{(Color online) Reciprocal-space reconstruction of 
a superconducting crystal of nominal \kfssc{} composition, with the
octagon of twinned peaks corresponding to \kfsfive{}. Reflections
are labeled with Miller indices of the $I4/mmm$ substructure.
Again, the (110) peak is present due to formation of the oxidized
phase after minor air exposure.
}
\label{fig:kspace-sc}
\end{figure}

Reciprocal-space reconstructions of a superconducting crystal 
from single-crystal x-ray
diffraction (Figure \ref{fig:kspace-sc}) show the supercell
reflections from vacancy-ordered \kfsfive{}. The single-crystal refined
composition is K$_{0.72(2)}$Fe$_{1.63(1)}$Se$_2$, but
the question of phase separation is
crucially important, since a distinct phase that induces superconductivity
may be present. 
\cite{ricci_nanoscale_2011,shermadini_superconducting_2012,lazarevic_vacancy_2012,liu_evolution_2012}
Our laboratory single-crystal diffractometer could
not resolve any new reflections that were not present in
nominal, non-superconducting \kfsfive{} crystals, so we performed 
high-resolution synchrotron powder diffraction to investigate the 
totality of phases present in these materials.

\begin{figure}
\centering\includegraphics[width=0.85\columnwidth]{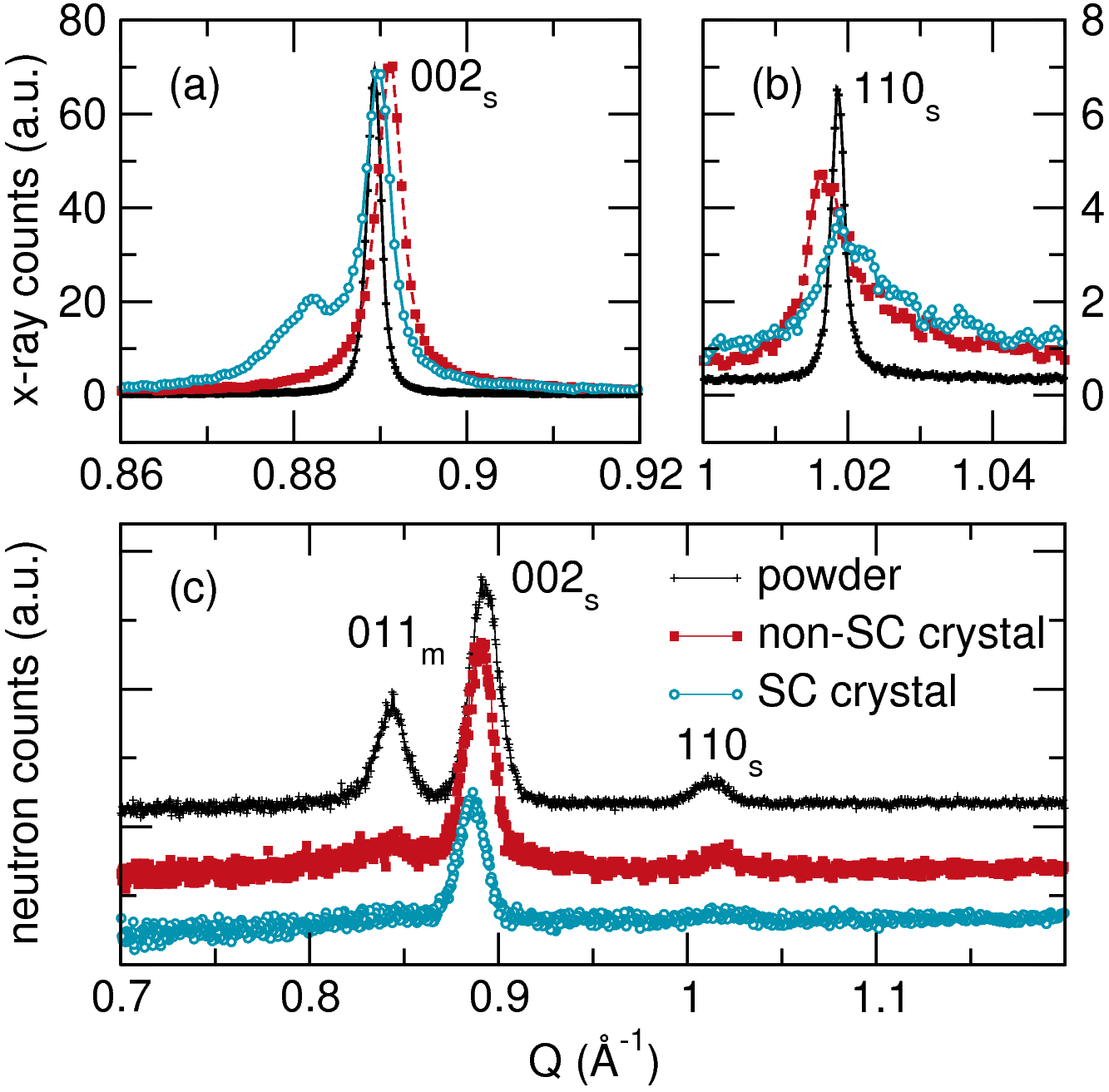} \\
\caption{(Color online) Comparison of diffraction peaks for powder \kfsfive{}
versus non-superconducting crystals (nominal \kfsfive{}) and superconducting crystals
(nominal \kfssc{}). 
All data were collected at 300 K.
In (a), the synchrotron diffraction peak (002) shows a split of the c-axis in
the SC sample. In (b), the magnified (110) peak shows significant disorder in 
the $ab$ plane in both crystals. In (c), the magnetic (011) reflection 
viewed by neutron diffraction from HIPD shows
strong magnetic ordering in the powder, weak magnetic ordering
in nominal \kfsfive{} crystals, and no magnetic ordering in 
superconducting nominal \kfssc{}.
}
\label{fig:peaks}
\end{figure}

A comparison of the Bragg peak splitting in superconducting
crystals (nominal \kfssc{})
and non-superconducting samples (nominal \kfsfive{}
crystals and powder) is shown in Figure
\ref{fig:peaks}. The superconducting crystals display a clear split of
the (002) reflection. This splitting is commonly seen when 
Rb, Cs, and K-containing single crystals
are characterized using simple Bragg-Brentano diffraction measurements,
\cite{luo_crystal_2011,wang_superconductivity_2011}
and most likely represents the metallic superconducting phase 
which we discuss in the next section.
There is no splitting in Figure \ref{fig:peaks}(a) for the
non-superconducting nominal \kfsfive{} crystal or powder.
This implies that phase separation is not an intrinsic
feature of pure \kfsfive{}. Rather, deviations from that stoichiometry are required
to drive phase separation. The (110) peak of the $I4/m$ vacancy-ordered
phase is compared in Figure \ref{fig:peaks}(b), and both single crystalline
samples are considerably broadened, with a long tail on the high-$Q$ side
of the peak, in the direction of $\beta$-FeSe which has
its (110) peak at $Q$ = 1.17 \AA$^{-1}$.

Stoichiometric deviation from pure \kfsfive{} leads to 
weakening of antiferromagnetic order. 
Neutron powder diffraction at 300 K in Figure \ref{fig:peaks}(c) shows
a strong $(011)$ magnetic peak at $Q$ = 0.84 \AA$^{-1}$, indicating strong
antiferromagnetic order in pure powder \kfsfive{}, which has $T_N$ = 559$^\circ$C.\cite{bao_novel_2011}
This peak is substantially weakened in the nominal \kfsfive{}
crystal, and has disappeared in the superconducting crystal. Indeed,
subtle changes in stoichiometry disrupt the magnetic ordering in the \kfsxy{}
lattice, and antiferromagnetism and superconductivity appear 
mutually exclusive.

\subsection{The superconducting metallic phase \kfsmetal{}}

\begin{figure}
\centering\includegraphics[width=0.85\columnwidth]{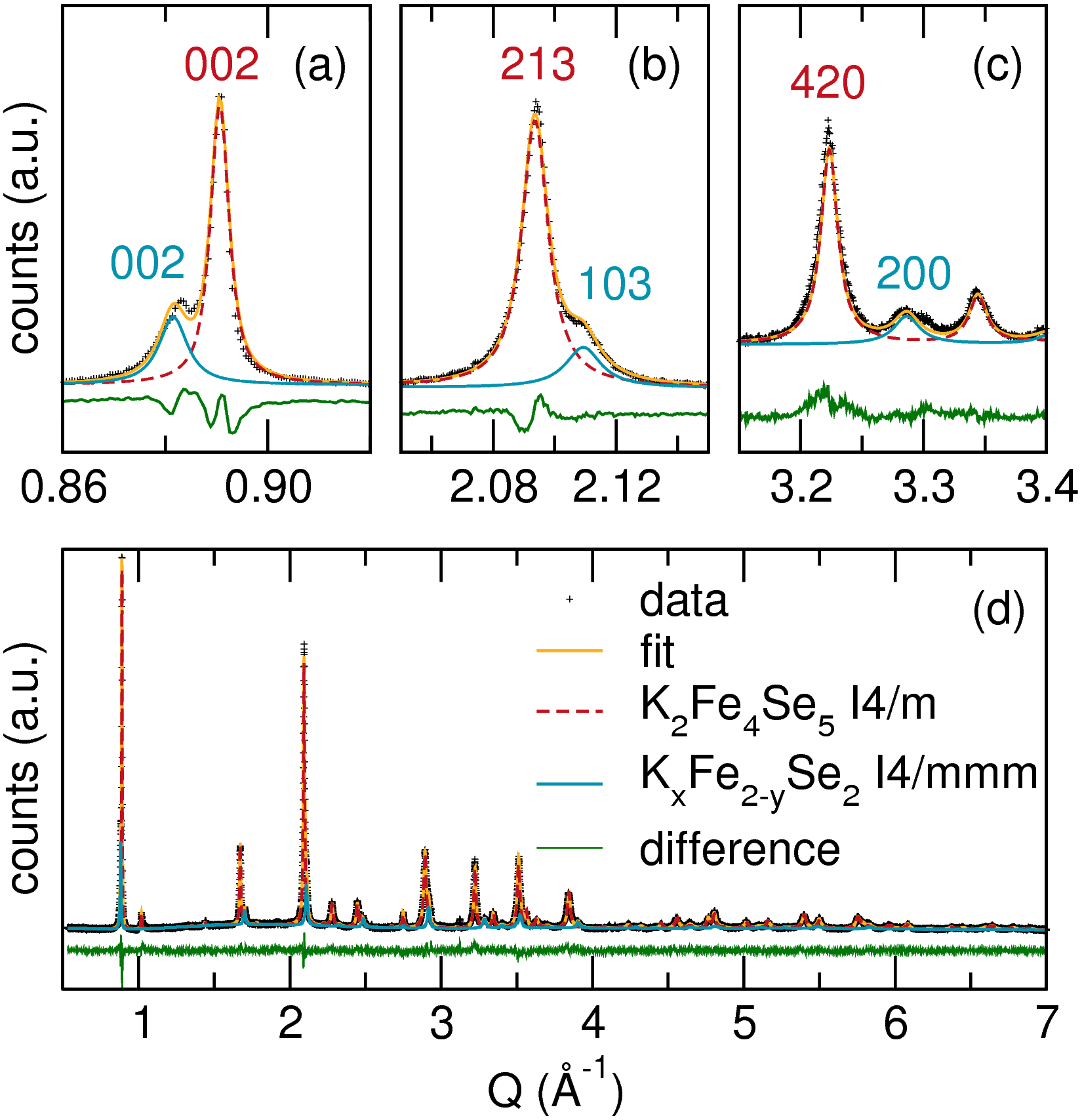} \\
\caption{(Color online)
Rietveld refinement to high-resolution synchrotron x-ray diffraction data
of a superconducting sample of nominal \kfssc{} composition displays
peak splitting corresponding to a distinct $I4/mmm$ phase at room
temperature. Selected
regions are expanded in (a,b,c) to show detail on equivalent pairs
of reflections. The labeled peaks would be coincident for both phases
in the absence of lattice distortions.
}
\label{fig:sc-rietveld-zoom}
\end{figure}

The 11-BM synchrotron x-ray data resolves splitting in
not only the  the (002) reflection, but an entirely separate $I4/mmm$ phase
that occurs in superconducting samples, shown in the insets of Figure
\ref{fig:sc-rietveld-zoom}(a,b,c). These extra peaks can be modeled using
a separate cell with disordered vacancies. For three separate
superconducting samples, this phase refines to a composition of 
\kfsmetal{} where $x$ = 0.38(2), 0.55(1), and 0.58(2),
with weight fractions of 13, 18, and 12\%.
Rietveld refinement results are summarized in Table \ref{tab:kfsmetal}.
Full details are given in the Supplemental Material.
Our compositions, along with those determined
by a lower-resolution diffraction study,\cite{lazarevic_vacancy_2012}
 and NMR measurements \cite{texier_nmr_2012}
all find evidence for the metallic minority phase to have nearly
full iron occupancy and K deficiency.
This phase must not display any K$^+$ vacancy ordering, as any 
superstructure peaks arising from $\sim$15\% of the sample would
be clearly visible in the single-crystal diffraction pattern (Figure \ref{fig:kspace-sc}). 
Recent high-temperature diffraction data have confirmed
that this phase is absorbed into \kfsfive{} above the
vacancy ordering temperature.\cite{liu_evolution_2012}

\begin{table}
\caption{\label{tab:kfsmetal} 
Rietveld refinement results for the superconducting metallic phase
\kfsmetal{} for three different samples. Sample numbers
correspond to points in Figure \ref{fig:latparam} and
to the full refinement details
and processing conditions given in the Supplemental Material.
}
\centering
\begin{tabular}{llllll}
\hline
\#   & $R_{wp}$ & Stoichiometry & wt\% & a (\AA) & c (\AA) \\
\hline 
3   & $R_{wp}$ & K$_{0.58(2)}$Fe$_{1.84(4)}$Se$_2$ & 12 & 3.83414(20) & 14.2360(12) \\
4   & $R_{wp}$ & K$_{0.55(1)}$Fe$_{2.00(2)}$Se$_2$ & 18 & 3.82803(23) & 14.2634(10) \\
6   & $R_{wp}$ & K$_{0.38(2)}$Fe$_{2.06(28)}$Se$_2$ & 13 & 3.82707(26) & 14.2658(15) \\
\hline
\end{tabular}
~\\
\end{table}

All three samples which exhibited phase separated \kfsmetal{} 
(3, 4, and 6 in Figure \ref{fig:latparam})
by synchrotron diffraction displayed
a diamagnetic response at $T_c$.
No semiconducting samples contained this minority $I4/mmm$ phase.
Two samples, (2 and 5 in 
Figure \ref{fig:latparam}) were superconducting but the
diffraction peaks were too broad to resolve the second phase due to quenching.
While samples with a small superconducting fraction can be made
reliably, creating homogeneous samples is a requirement for
understanding the mechanisms of superconductivity in these samples,
for example by photoemission spectroscopy or inelastic neutron scattering.
To that end, we have begun to map the available phases in the \kfsxy{}
system and probe their stability.

The previous claim that KFe$_2$Se$_2$ is the superconducting phase 
seems implausible since this formula requires half to the Fe atoms 
to be in the 1+ state and tetrahedrally coordinated by Se. Such a 
state is unlikely to be stable since it requires excessive 
negative charge on a large fraction of the Fe atoms and is 
unprecedented in the literature. On the other hand, a \kfsmetal{} 
formulation with $x \sim 0.5$ would require only a quarter of Fe 
atoms to be in a 1+ state and in this case the extra negative charge 
may be delocalized over a broad conduction band.

\subsection{Changes in $I4/m$ structure of \kfsfive{} seen by 
Rietveld refinement}

\begin{figure}
\centering\includegraphics[width=0.99\columnwidth]{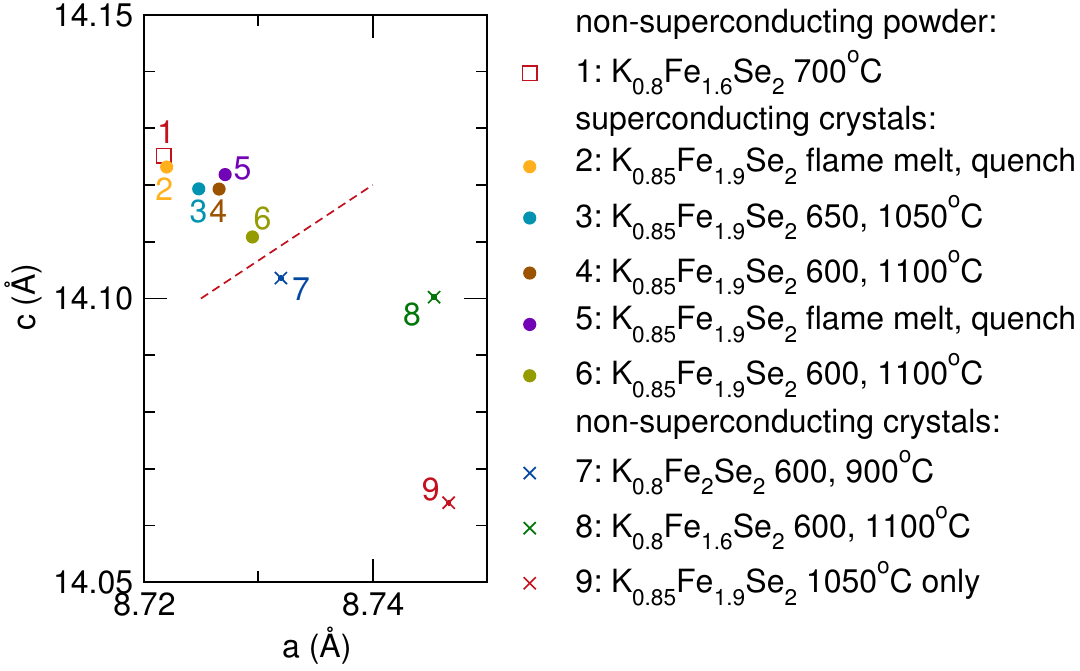} \\
\caption{(Color online) Lattice parameters for the bulk \kfsxy{} phase
with $I4/m$ structure obtained from Rietveld refinement of high-resolution 
synchrotron
diffraction data display a narrow range of $c$-axis for superconducting
samples, but only in crystals grown from the melt. Temperatures of
pre-reaction and subsequent crystal growth are shown for each sample.
Refined parameters and detailed synthesis conditions
are given in the Supplemental Material.
}
\label{fig:latparam}
\end{figure}

In all our samples, regardless of superconductivity, the $I4/m$ \kfsfive{}
phase is present. We have searched via Rietveld refinement
for systematic changes
in the $I4/m$ phase that might be associated with the onset
of superconductivity.

Lattice parameters for the $I4/m$ phase are given in Figure \ref{fig:latparam}.
Sample 1 is a pure powder (non-superconducting) of \kfsfive.
The cluster of superconducting samples all have an $a$-axis
smaller than 8.73 \AA{} and a $c$-axis larger
than 14.11 \AA{}, distinct from the non-superconducting crystals and
separated by a dashed line.
However these lattice parameters are not a structural
trigger of superconductivity, since the insulating powder sample falls
in the superconducting range.
Instead we assert that the superconducting samples contain a majority
$I4/m$ component that is near the edge of its compositional range,
and so are predisposed to containing the minority \kfsmetal{} 
superconducting phase.

\begin{figure}
\centering\includegraphics[width=0.85\columnwidth]{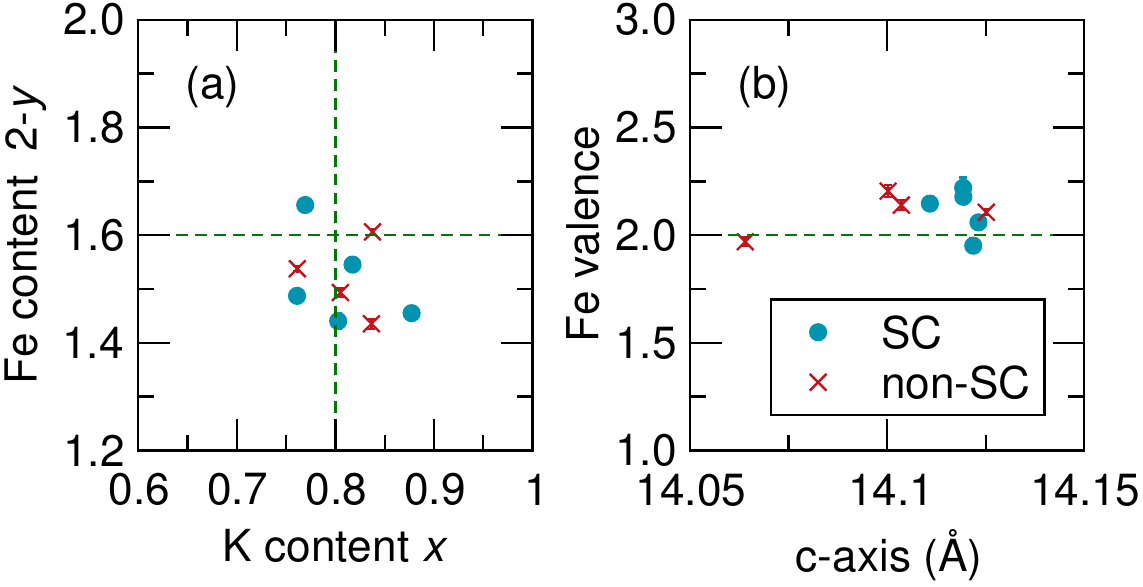} \\
\caption{(Color online) K and Fe content from Rietveld
refinements of the majority $I4/m$ phases in superconducting (SC) and 
non-superconducting samples
show no clear distinction between the two groups. Fe
tends to be slightly deficient in superconducting samples.
 Fe valence derived from Rietveld-refined
stoichiometry shows a strong tendency for majority Fe$^{2+}$ in the
$I4/m$ phase.
}
\label{fig:stoich}
\end{figure}

The stoichiometries of all $I4/m$ \kfsxy{} phases from Rietveld refinements
are shown in Figure \ref{fig:stoich}(a).
The superconducting samples are tightly clustered in composition space,
but there is no distinction between them and the non-superconducting
samples. K contents are near nominal values, 
while Fe tends to be deficient, around K$_x$Fe$_{1.5}$Se$_2$.
An approximate calculation of Fe valence using these refined
stoichiometries is shown in Figure \ref{fig:stoich}(b).
The non-superconducting and superconducting samples are
both clustered around Fe$^{2+}$.

A clear division was seen in lattice parameters (Figure
\ref{fig:latparam}) for superconducting and non-superconducting
samples, but not in the refined stoichiometry (Figure \ref{fig:stoich}). As a
result, the lattice parameters may be a more exact probe
of the response of the \kfsfive{}-type $I4/m$ lattice
to stoichiometry, and further work should be done
to explain how the lattice parameters change with K and
Fe content, and their relation to phase separation,
which is now believed to be necessary for
superconductivity. \cite{torchetti_se_2011,ksenofontov_phase_2011,ma_local_2011}
We present preliminary work on this subject
in the next section.

\subsection{Comparing related phases in the \kfsxy{} series: 
$\beta$-FeSe, \kfsmetal{}, \kfsfive{}, and KFe$_{1.6}$Se$_2$}

\begin{figure}
\centering\includegraphics[width=0.85\columnwidth]{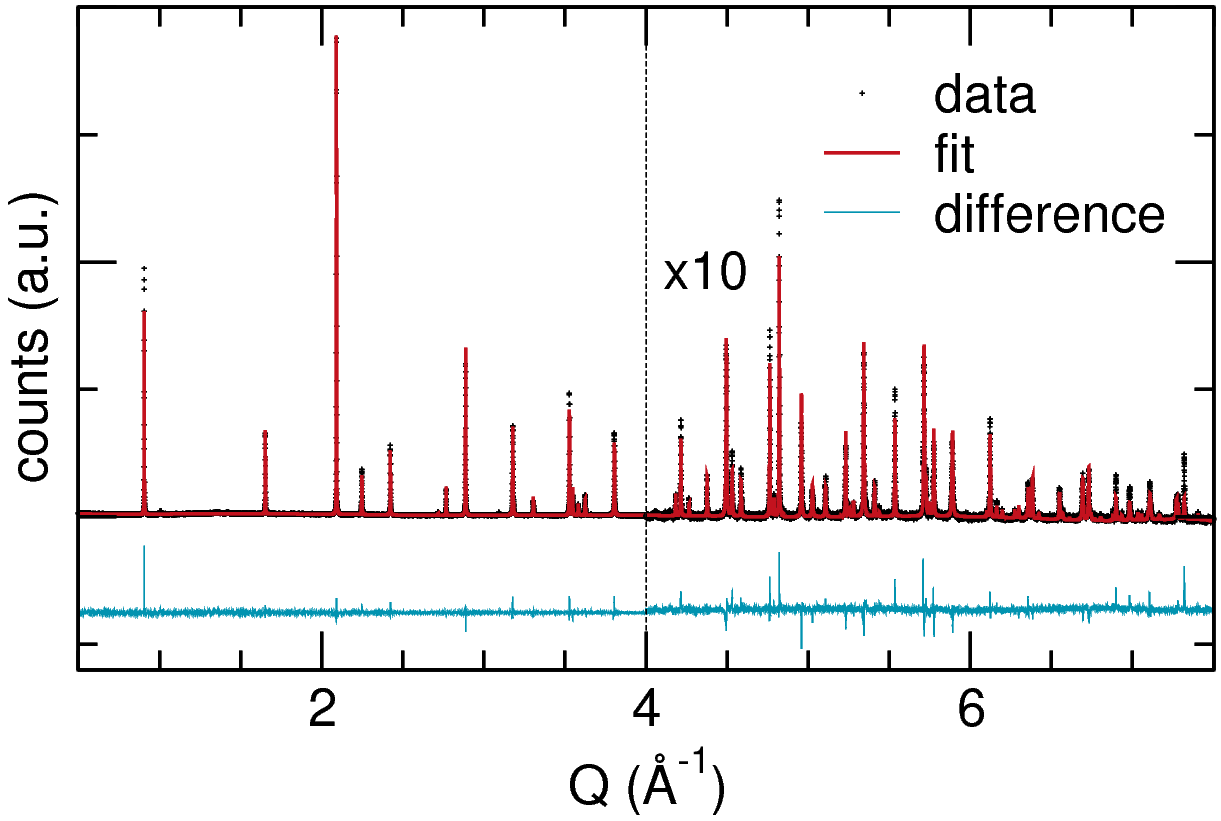} \\
\caption{(Color online)
Synchrotron x-ray diffraction Rietveld
refinement of a vacancy-disordered, non-superconducting
K$_{0.959(4)}$Fe$_{1.606(6)}$Se$_2$ powder sample
with fully-occupied K sites. Refinement results are given in
the Supplemental Material.
}
\label{fig:rietveld-234}
\end{figure}

\begin{figure}
\centering\includegraphics[width=0.85\columnwidth]{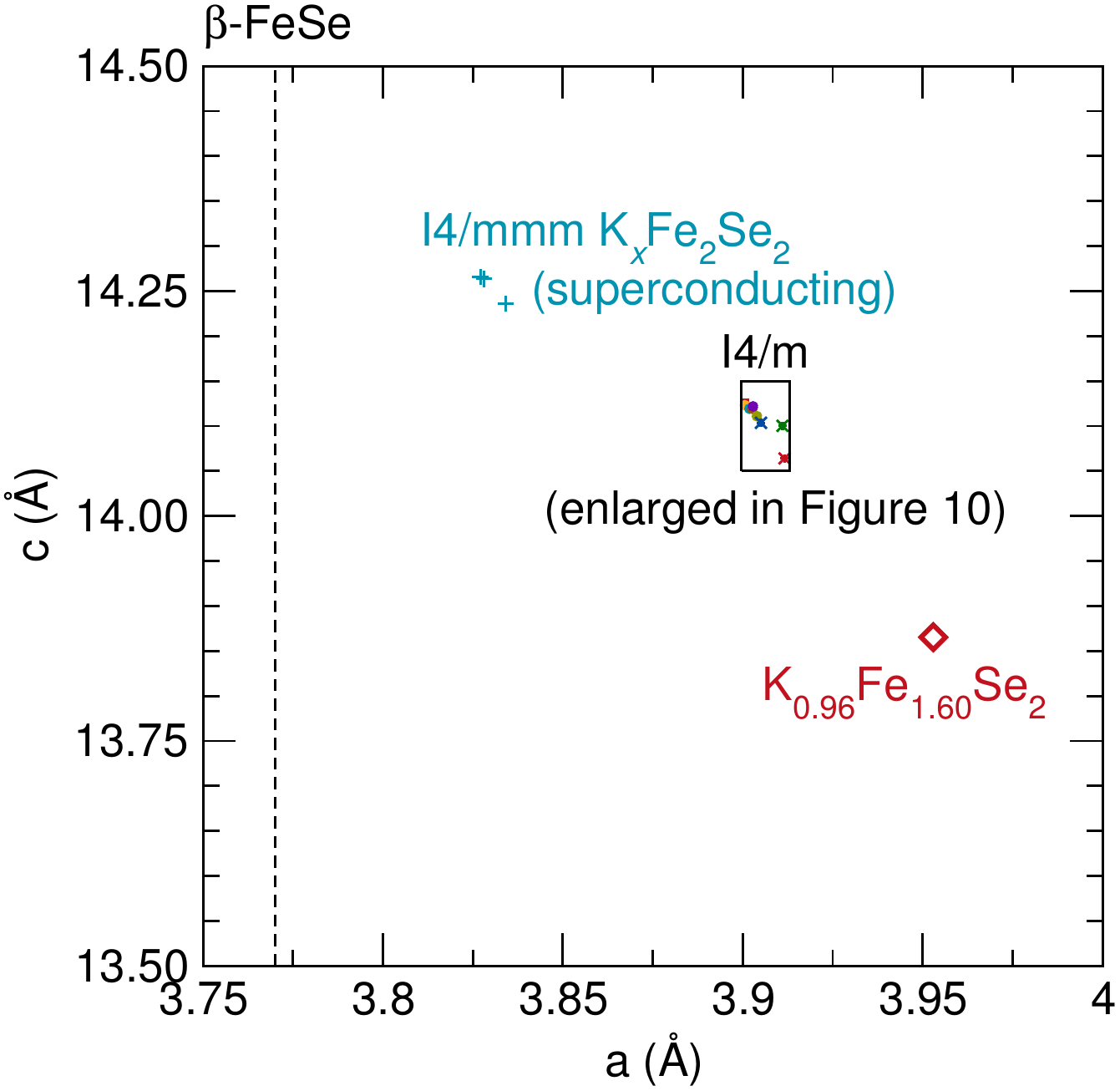} \\
\caption{(Color online) Trends of $a$ and $c$ lattice parameters
at 300 K across the \kfsxy{} phase space, as determined by
Rietveld refinement to high-resolution synchrotron diffraction data.
The trend implies that the minority phase in superconducting
samples is very K-deficient, and distinct from the 
stability regions of pure $\beta$-FeSe or $I4/m$ \kfsfive{}.
}
\label{fig:ac-trend}
\end{figure}

Hypothetically, the \kfsxy{} phase space could contain a plethora
of homologous (K$_2$Se)(FeSe)$_n$ phases containing strictly Fe$^{2+}$,
from $n=3$ K$_2$Fe$_3$Se$_4$, where the K
layer is filled, to $n=\infty$ corresponding to $\beta$-FeSe.
Our attempts to produce phases with higher $n$ (K$_2$Fe$_5$Se$_6$, K$_2$Fe$_6$Se$_7$,
etc.) by solid state reactions simply
led to \kfsfive{} + $\beta$-FeSe. Reactions
with the nominal composition K$_2$Fe$_3$Se$_4$ gave
a pure compound, and upon synchrotron x-ray diffraction the refined
occupancy was found to be K$_{0.959(4)}$Fe$_{1.606(6)}$Se$_2$, with excess K and Se likely
precipitating as amorphous K$_2$Se$_4$.\cite{sangster_k-se_1997}
There are no superstructure
peaks in this compound, indicating that the Fe vacancies are truly
disordered and the symmetry remains $I4/mmm$.  The fit from Rietveld
refinement is shown in Figure \ref{fig:rietveld-234}, and  
results are tabulated in the Supplemental Material.
The isostructural phase TlFe$_{1.6}$Se$_2$ exhibits  multiple magnetic transitions
at low temperatures,
\cite{may_spin_2012,sales_unusual_2011}
so further investigation is warranted.
We did not detect any superconducting diamagnetic response in KFe$_{1.6}$Se$_2$ 
down to 2 K.

The KFe$_{1.6}$Se$_2$ phase represents a third distinct
phase in the \kfsxy{} system, in addition to \kfsfive{}
and \kfsmetal{}. The lattice parameters of all
these phases are shown in Figure \ref{fig:ac-trend},
(with the $I4/m$ phase normalized by $\sqrt 5$).
This diagram provides a full view of known phases in
the \kfsxy{} system, from full K occupancy in KFe$_{1.6}$Se$_2$
to empty interlayer space in $\beta$-FeSe. A trend 
of decreasing $a$ with increasing $c$ is evident, 
likely driven by weak van der Waals interlayer
forces yielding to stronger ionic bonding as K$^+$
is inserted. Simultaneous carrier donation from K$^+$ into
the FeSe layers leads to weaker Fe-Se bonding and an
increase in intralayer distances ($a$ lattice parameters).
From this plot, the separation between
$I4/m$ \kfsfive{} phases and the superconducting minority
phases is shown to be quite significant.
We discuss implications for the synthesis of this phase subsequently.
It remains to be seen if there is a solid solution between
\kfsfive{} and KFe$_{1.6}$Se$_2$.

\subsection{Superconducting composite of \kfsfive{} and \kfsmetal{}: Magnetometry and heat capacity}

\begin{figure}
\centering\includegraphics[width=0.85\columnwidth]{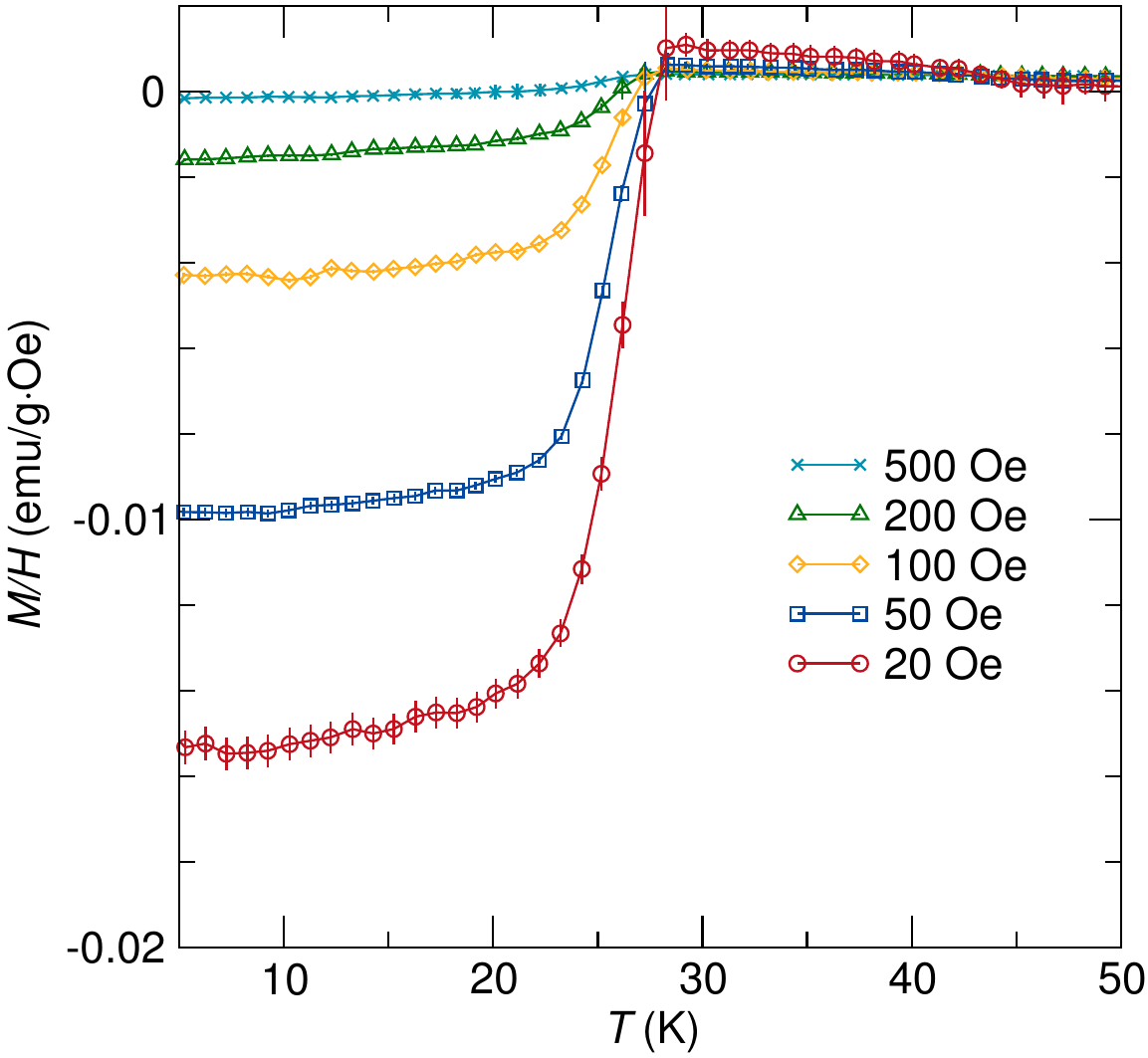} \\
\caption{(Color online) Magnetization of a superconducting
sample of nominal \kfssc{} composition.
}
\label{fig:acms-sc}
\end{figure}

DC magnetometry of a superconducting sample (sample 6)
is shown in Figure \ref{fig:acms-sc}, with $T_c$ = 28 K.
Such measurements are unfortunately not a viable way to probe the 
superconducting volume fraction.  If the  fraction is small but 
pervades the entire sample, as in a net-like 
model,\cite{shen_intrinsic_2011} then magnetometry would give an 
inflated view of the volume fraction. For this reason, we performed 
heat capacity measurements on samples that had already been confirmed
to be superconducting by magnetometry.

\begin{figure}
\centering\includegraphics[width=0.85\columnwidth]{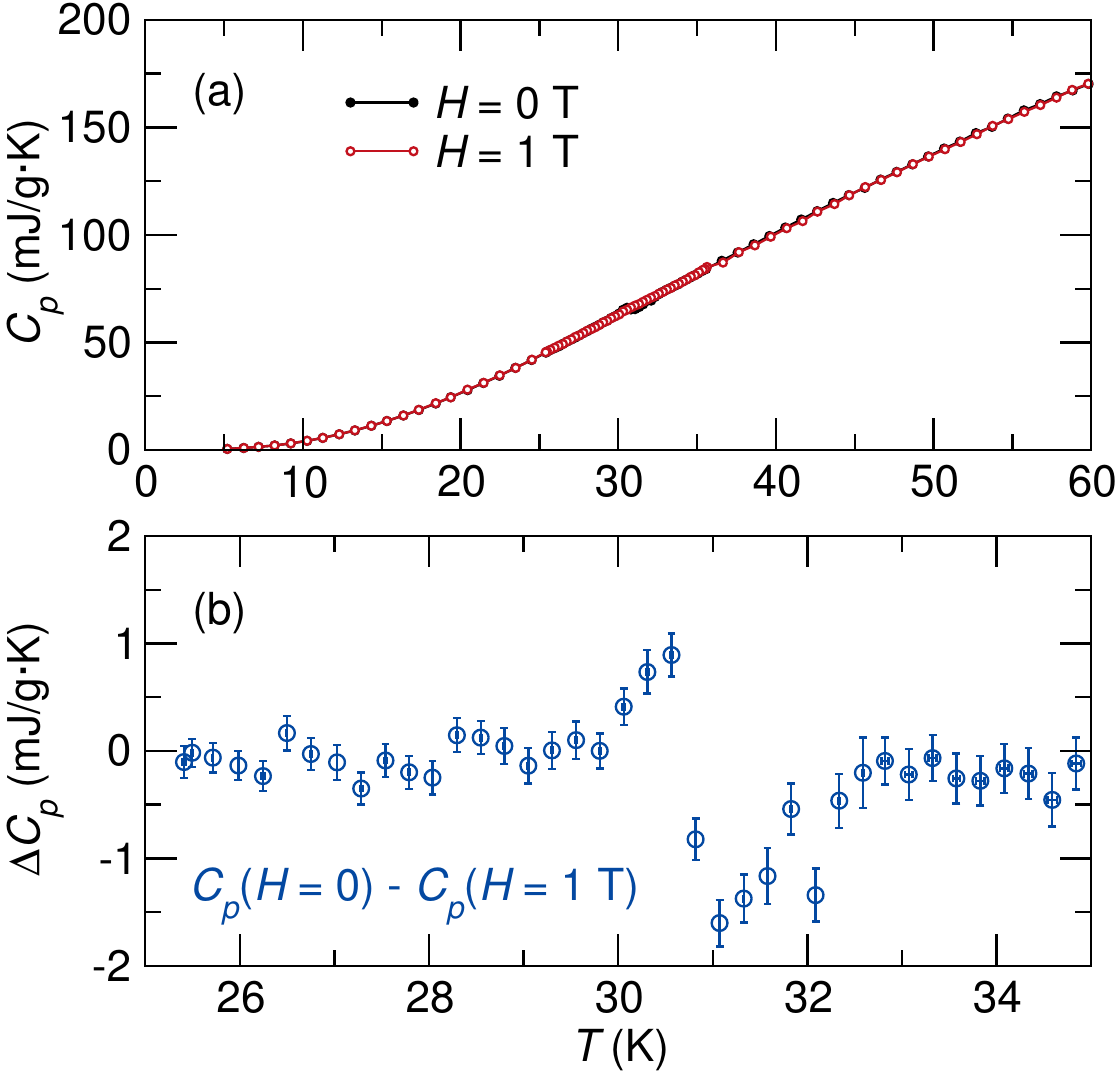} \\
\caption{(Color online) Heat capacity of a superconducting
crystal shows a very small anomaly at $T_c$. This feature
is magnified in (b) by subtracting the $H$ = 1 T measurement
from the zero-field measurement.
}
\label{fig:heatcapacity}
\end{figure}

Heat capacity measurements provide an excellent way to confirm 
bulk superconductivity, although the precise volume fraction would
depend on a known model for the entropy release at $T_c$.
Studies on YBa$_2$Cu$_3$O$_{7-\delta}$ and $\beta$-FeSe
have shown clear signatures of entropy
release ($\Delta C_p \approx 6.9$ and 3 mJ/gK, respectively)
across $T_c$. \cite{liang_growth_1992,mcqueen_extreme_2009}
We measured a single crystal (sample 6) with a strong zoom across $T_c$
and the measurement is seen in Figure \ref{fig:heatcapacity}. 
The inset
in Figure \ref{fig:heatcapacity}(b) shows the difference between heat capacity
measured at zero field and $H=1$ T. The anomaly at $T$ = 31 K is approximately
2 mJ/gK, which is comparable to $\beta$-FeSe, even though the 
fraction of \kfsmetal{} phase is only $\sim$15\% by weight.
The small peak in this data confirms the minor phase fraction of superconducting
\kfsmetal{} seen in powder diffraction patterns and magnetic susceptibility. Further
evidence for this two-phase coexistence is seen in resistivity measurements.

\subsection{Resistivity: Metal-insulator crossover implies two-phase coexistence}

\begin{figure}
\centering\includegraphics[width=0.9\columnwidth]{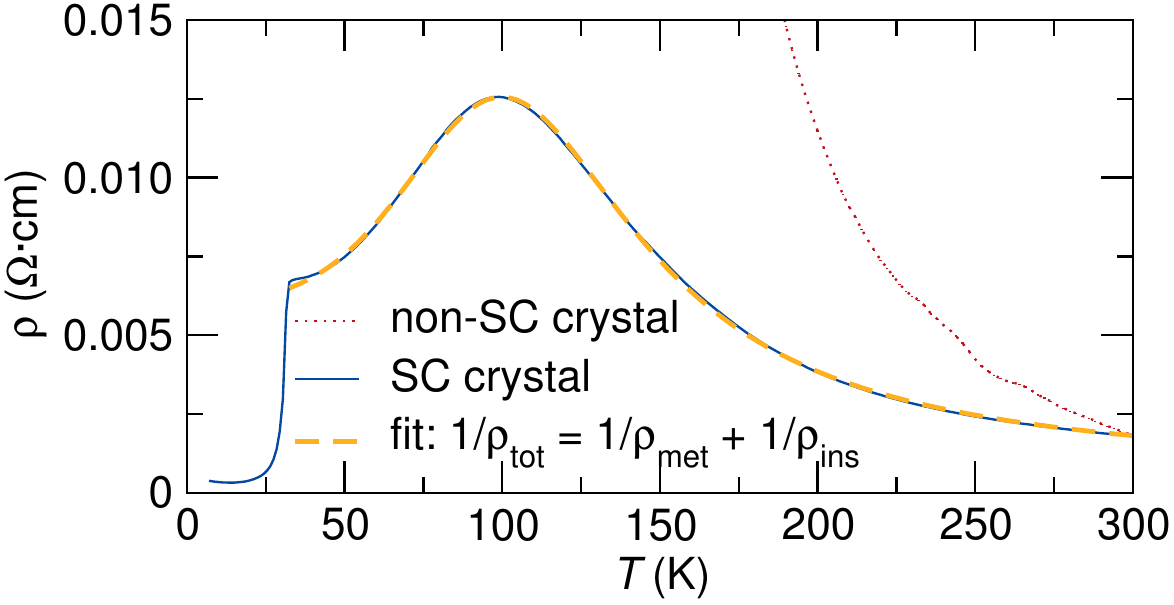} \\
\caption{(Color online) Resistivity of non-superconducting \kfsfive{}
crystal and a nominal \kfssc{} superconducting crystal. The 
\kfsfive{} is an insulator, but the superconductor behavior
can be fit as a metallic and insulating composite over the 
full temperature range.
}
\label{fig:resistivity}
\end{figure}

Resistivity ($\rho$) measurements from superconducting and non-superconducting
crystals are shown in Figure \ref{fig:resistivity}.
The $\rho$ drops to zero at $T_c$ = 31 K, in agreement
with our magnetization and heat capacity measurements.
There is a hump in the resistivity
around 100 K, which was seen in many studies, including the initial report by Guo,
et al
\cite{guo_superconductivity_2010,ying_superconductivity_2011,yan_electronic_2012} 
and attributed
to a metal-insulator transition.\cite{lei_phase_2011} However, given
the phase separation between \kfsfive{} and \kfsmetal{},
it is more likely that the metallic and insulating
phases are \emph{always} present. Thus the full resistivity 
range can be fit using a model of two percolating phases that 
act as resistors in parallel,
one with metallic Bloch-Gr\"{u}neisen temperature dependence $\rho_{metal}$
and the other with Boltzmann-type insulating temperature dependence
$\rho_{ins}$:

\begin{equation}
1/\rho_{total} = 1/\rho_{metal} + 1/\rho_{semi}
\end{equation}
\begin{equation}
\rho_{metal}(T) = \rho(0) + AT^n
\end{equation}
\begin{equation}
\rho_{semi}(T) = \rho_0 e^\frac{E_g}{2kT}
\end{equation}

Where $\rho(0)$, $A$, and $\rho_0$ are all constants that depend on
phase fractions and geometry in this case.
This fit (dashed in Figure \ref{fig:resistivity})
is excellent and gives $n$ = 2.78, and insulator activation
energy of $E_g$ = 83 meV.
Changes in the position of the hump
can be accomplished by simply changing the relative 
volume fractions of these two phases.
The metallic phase is not iron since it is present in such
small amounts ($\leq 2\%$ by weight by synchrotron powder diffraction).
Furthermore, muon spin rotation and scanning probe
measurements indicate that the superconducting phase is metallic above $T_c$.
\cite{torchetti_se_2011,charnukha_nanoscale_2012,li_phase_2012}
This resistivity maximum provides further confirmation that the minority
$I4/mmm$ phase is the cause of superconductivity,
and further work should be conducted to optimize its synthesis.

\subsection{Implications for synthesis}

Pure, bulk superconducting samples of \kfsxy{} remain elusive,
but careful structural studies  can explain
why this phase is difficult to
synthesize.
First, it is surprising that \kfsmetal{} forms from solid state reactions
because all known alkali iron chalcogenides have Fe valence nearly 
2+ or 3+. We attempted to intercalate
K into $\beta$-FeSe by vapor transport in a sealed tube at 300$^\circ$C.
However this reaction only resulted in the formation of 
K$_2$Se and metallic Fe, and no increase in $T_c$ above 8 K.
Why then does Fe-rich phase \kfsmetal{} form during heat treatment of
\kfsxy{}?

We propose that the metallic superconducting fraction
precipitates upon cooling through
the Fe vacancy ordering
temperature at $\sim$540 K,\cite{liu_evolution_2012} but only in cases where the crystal
size is large enough for lattice strain to prevent escape of 
supersaturated Fe from the \kfsxy{} structure.
Formation of a coherent intergrowth
of this $I4/mmm$ phase is 
supported by recent evidence from electron microscopy and
muon spin rotation.\cite{ryan_fe_2011,charnukha_nanoscale_2012}
We have not observed superconductivity in polycrystalline
powder samples, indicating lattice strain may be a key factor. If $T_c$ is eventually
observed in powders, it would mean that the Fe supersaturation
in the \kfsxy{} structure at high temperatures is the only
prerequisite for formation of \kfsmetal{}.

Only a small amount of Fe excess can be incorporated
in the \kfsxy{} structure at high temperatures. This places a limit on 
the volume of minority $I4/mmm$ phase that will precipitate
when cooling through the vacancy ordering temperature. The
separation between the maximum Fe solubility at 
high temperatures and 80\% Fe occupancy (in \kfsfive{})
determines the amount of \kfsmetal{} that can form.
This explains why superconducting
samples show $I4/m$ lattice parameters on the 
edge of the \kfsfive{} stability region in 
Figure \ref{fig:latparam}, and
why the heat capacity measurements and powder diffraction
both find a small volume fraction of superconducting phase.

Solid-state routes toward single-phase, superconducting \kfsmetal{}
will require an understanding of, and control over, the delicate
temperature-composition space in the region between \kfsfive{}
and $\beta$-FeSe. In-situ experiments (diffraction,
calorimetry, or vibrational spectroscopy) that investigate 
the limit of Fe solubility in \kfsxy{} around and above the vacancy
ordering temperature may prove invaluable. Quenching from
above this temperature has shown to increase the 
sharpness of the superconducting transition,
\cite{han_metastable_2012,ozaki_one-step_2012} and 
understanding the kinetics of this transition may provide insight
into stabilizing Fe-rich phases. Topotactic reactions
at low temperatures, such as those conducted in liquid ammonia,
seem to have the ability to intercalate $\beta$-FeSe without significant expulsion
of Fe,
\cite{burrard-lucas_enhancement_2012,ying_observation_2012,krzton-maziopa_synthesis_2012}
while oxidative deintercalation as was performed on KNi$_2$Se$_2$
may approach \kfsmetal{} by removal of K$^+$.\cite{neilson_bonding_2012}

Expanding the available $I4/mmm$ composition space by doping may
provide new routes to stabilize phases similar to \kfsmetal{}.
The response of ThCr$_2$Si$_2$ structures with substitution
of Se$^{2-}$ for As$^{3-}$ has not been systematically investigated.
Only the solid solution K$_x$Fe$_{2-y}$(Se,S)$_2$ has been
investigated (albeit without a description of subtle phase
separation). \cite{lei_phase_2011}
Even simple phase equilibria studies, such as the evolution of 
phases across nominal K$_x$Fe$_2$Se$_2$ ($0 \leq x \leq 1$)
from room temperature to $\sim$1250 K remain unknown.

\section{Conclusions} 

The stable phase close to superconducting stoichiometry,
vacancy-ordered \kfsfive{} phase can be made pure 
by a solid state powder reaction. We find
no evidence that this $I4/m$ phase can be doped or substituted to become
superconducting. As a result, 
high-resolution diffraction experiments are needed to detect the presence
of additional phases.

The metallic minority phase \kfsmetal{} with $I4/mmm$ symmetry appears
only in samples that exhibit superconductivity, as judged by 
a diamagnetic response around $T_c$ = 30 K. This phase
does not exhibit any vacancy ordering.
It only occurs in large crystals
of \kfsxy{} grown from the melt, so the excess Fe is likely
trapped by lattice strain, forming a coherent intergrowth with a volume
fraction that is limited by the solubility of excess Fe above
the vacancy ordering temperature of \kfsfive{}. This model of
phase separation is supported
by our resistivity measurements, which indicate a percolative
composite of an insulator and metal, which is supported by
local NMR and muon spin resonance probes and electron microscopy studies.
\cite{torchetti_se_2011,texier_nmr_2012,charnukha_nanoscale_2012,liu_evolution_2012}

We identified an oxidized phase K$_{0.51(5)}$Fe$_{0.70(2)}$Se
as the cause of (010) reflections
in the single-crystal diffraction pattern that are forbidden by
$I$-centered symmetry. This phase has a FeSe interlayer
spacing of 9 \AA{}, which is highly expanded versus the 7 \AA{} spacing of
\kfsfive{}, due to buckling of the K layer after oxidation of Fe and
loss of Se. This phase forms in the PbClF structure, similar to NaFeAs.
It is not relevant to superconducting behavior,
and sufficient care must be taken to prevent exposure of \kfsxy{}
samples to moisture.

Yet another phase, KFe$_{1.6}$Se$_2$ was identified to form
with disordered vacancies ($I4/mmm$) and pure polycrystalline powders
were obtained by solid state reaction. This phase was produced
when we attempted to synthesize the hypothetical ordered
compound K$_2$Fe$_3$Se$_4$ in the homologous series (K$_2$Se)(FeSe)$_n$.
The response of the \kfsxy{} lattice as stoichiometry
is varied from  KFe$_{1.6}$Se$_2$ to \kfsfive{}, \kfsmetal{},
and $\beta$-FeSe may prove to be a valuable probe of phase
equilibria and electrical response in these systems, especially
because the Rietveld-refined K/Fe stoichiometry does not provide a definitive
picture of the divide between superconducting and non-superconducting
samples.

Further investigations of superconducting \kfsmetal{} 
must embrace the fact that these phases are unstable and
heterogeneous. More informed synthesis should be pursued
by investigating the high-temperature phase relations
in these systems, and by understanding the kinetic
processes occurring when the superconducting minority
phase separates from related \kfsfive{}.

Finally, the insights obtained from this work call for detailed 
transmission electron microscopy studies to understand the 
superconducting/semiconducting interface and assess the nature of 
strain and defects associated with it. Clearly, bulk phase separation 
can form such composite structures. Phase separation 
can proceed by nucleation and growth or spinodal decomposition. 
The dividing line between them depends crtically on the strain that 
develops in the system during phase separation. According to our 
studies, \kfsfive{} and \kfsmetal{} have a lattice mismatch of 1-2\%, 
leading to considerable strains. Our present results call for first 
principles studies of the thermodynamics of incoherent and coherent 
phase separation in the \kfsfive{}/\kfsmetal{} systems to calculate 
strain energies and mixing energies.

\section{Acknowledgments}
Work at Argonne National Laboratory is supported by UChicago Argonne,
a U.S. DOE Office of Science Laboratory, operated under Contract 
No. DE-AC02-06CH11357.
This work utilized the HIPD instrument at the Los Alamos
Neutron Science Center, funded by the DOE Ofﬁce of Basic
Energy Sciences and operated by Los Alamos National Security
LLC under Contract No. DE-AC52-06NA25396.

\bibliography{kfese}

\end{document}